\documentclass[sort&compress
    ,final            
  ]
  {aipproc}

\usepackage{sidecap,graphicx,boxedminipage,epsfig,wrapfig,floatflt,shadow}
\layoutstyle{8x11double}
\pdfoutput=1

\begin{document}

\title{Quantum Interactive Learning Tutorials}

\author{Chandralekha Singh}{
  address={Department of Physics, University of Pittsburgh, Pittsburgh, PA, 15213}
}

\begin{abstract}
We discuss the development and evaluation of Quantum Interactive Learning Tutorials (QuILTs), 
suitable for one or two-semester undergraduate quantum mechanics courses.
QuILTs are based upon investigation of student difficulties in learning quantum physics.
They exploit computer-based visualization tools and  help students build links between
the formal and conceptual aspects of quantum physics without compromising the technical content.
They can be used both as supplements to lectures in the classroom or as a self-study tool.
\end{abstract}
\maketitle

\section{Introduction}

Quantum physics is a technically difficult and abstract subject.~\cite{griffiths} 
The subject matter makes instruction quite challenging and students perpetually struggle to master basic 
concepts.~\cite{galvez,zollman,styer,narst,theme,my1,phystoday,my2,my3,my4,my5,fischler}
Here I discuss the development and evaluation of Quantum Interactive Learning Tutorials (QuILTs) that help 
advanced undergraduate students learn quantum mechanics.
QuILTs are designed to create an active learning environment where students have an opportunity to 
confront their misconceptions, interpret the material learned, 
draw qualitative inferences from quantitative tools learned in quantum mechanics and
build links between new material and prior knowledge. 
They are designed to be easy to implement regardless of the lecturer's teaching style. 

A unique aspect of QuILTs is that they are research-based, targeting specific 
difficulties and misconceptions students have in learning various concepts in 
quantum physics.~\cite{zollman,styer,narst,theme,my1,phystoday,my2,my3,my4,my5,fischler} 
They often employ computer-based visualization tools~\cite{mario,pqp,others,mcintyre,zollman1,styer_cups} to help students 
build physical intuition about 
quantum processes and keep students consistently engaged in the learning process by asking them
to predict what should happen in a particular situation, and then providing appropriate feedback.
They attempt to bridge the gap between the abstract quantitative formalism of quantum mechanics and the qualitative 
understanding necessary to explain and predict diverse physical phenomena. They
can be used in class by the instructors once or twice a week as supplements to lectures or outside of the class
as homework or as self-study tool by students. 

\section{Details of the QuILTs}

The QuILTs  use a learning cycle approach~\cite{wiki} in which students' engage in the topic via examples
that focus their attention, explore the topic through facilitated questioning and observation, explain what they have
learned with instructor facilitating further discussion to help refine students' understanding and
extend what they have learned by applying the
same concepts in different contexts. 
The guidance provided by the tutorials is decreased gradually and students
assume more responsibility in order to develop self-reliance. 

In addition to the main tutorial, QuILTs often have a ``warm-up" component and a tutorial ``homework".
Students work on the ``warm-up" component of a QuILT at home before working on the main tutorials in class. 
These warm-ups typically review the prior knowledge necessary for optimizing the benefits of the main
tutorial related to that topic. The ``tutorial homework" associated with a QuILT can be given
as part of their homework to reinforce concepts after students have worked on the main tutorial.
The tutorial homework helps students apply the topic of a particular QuILT to many specific situations 
to learn about its applicability in diverse cases and learn to generalize the concept appropriately.

We design a pre-test and post-test to accompany each QuILT.
The pre-test assesses students' initial knowledge before they have worked on the corresponding QuILT, but typically
after lecture on relevant concepts. 
The QuILT, together with the preceding pre-test, often make students' difficulties related to relevant
concepts clear not only to the instructors but also to students themselves. 
The pre-test can also have motivational benefits and can
help students better focus on the concepts covered in the QuILT that follows it.
Pre-/post-test performances are also useful for refining and modifying the QuILT.

An integral component of the QuILTs is the adaptation of visualization tools 
for helping students develop physical intuition about quantum processes. 
A visualization tool can be made much more pedagogically effective if it is embedded in a learning environment such as QuILT.
A simulation, preceded by a prediction and followed by questions, can help students reflect upon what they 
visualized. Such reflection can be useful for understanding and remembering concepts.
They can also be invaluable in helping students better understand the differences between classical and quantum concepts.

We have adapted simulations from a number of sources as appropriate~\cite{mario,others,mcintyre,zollman1,styer_cups} 
including the open source JAVA simulations developed by Belloni and Christian.~\cite{mario}
Some of the QuILTs, e.g., the QuILT on double-slit experiment which uses simulations developed by Klaus 
Muthsam,~\cite{others} are also
appropriate for modern physics courses. The double-slit QuILT uses simulations to teach students about 
the wave nature of particles manifested via the double slit experiment with single particles,
the importance of the phase of the probability amplitude for the occurrence of interference pattern and
the connection between having information about which slit a ``particle" went through (``which-path" information)
and the loss of interference pattern.

For the QuILTs based on simulations, students must first make predictions about what and why they
expect a certain thing to happen in a particular situation before exploring the relevant concepts with the simulations.
For example, students can learn about the stationary states of a single particle in 
various types of potential wells. Students can change the model parameters and learn how those parameters affect stationary states and
the probability of finding the electron at a particular position. They can also take various linear combinations of stationary states
to learn how the probability of finding the electron at a particular position is affected.
They can calculate and compare the expectation values of various operators in different states for a given potential.
They can also better appreciate why classical physics may be a good approximation to quantum physics under certain conditions. 
Students can also develop intuition about the differences between bound states and scattering states by using visual simulations. 
Guided visualization tools can also help students understand the changes that take
place when a system containing one particle is extended to many particles.~\cite{styer_cups}

Similar to the development of tutorials for introductory and modern physics,~\cite{lillian,redish}
the development of each QuILT goes through a cyclical iterative process.
Preliminary tutorials are developed based upon common difficulties in learning a particular 
topic,~\cite{zollman,styer,narst,theme,my1,phystoday,fischler}
and how that topic fits within the overall structure of quantum mechanics.
The preliminary tutorials are implemented in one-on-one interviews with student volunteers, and modifications are
made. These modifications are essential for making the QuILTs effective.
After such one-on-one implementation with at least half a dozen students, tutorials are tested and evaluated
in classroom settings and refined further.

Working through QuILTs in groups is an effective way 
of learning because formulating and articulating thoughts can provide students with an opportunity to solidify concepts and
benefit from one another's strengths. It can also provide an opportunity to monitor their own learning because mutual
discussions can help students rectify their knowledge deficiencies. 
Students typically finish a QuILT at home if they cannot finish it in class and
take the post-test associated with it individually in the following class for which no help is provided.

\section{Case Studies}

Below, we briefly discuss case studies related to the development and evaluation of three QuILTs on time development
of wave function, uncertainty principle, and Mach-Zehnder interferometer. The development of each QuILT starts with
a careful analysis of the difficulties students have in learning related concepts. After the preliminary development
of the tutorials and the pre-/post-tests associated with them, we conduct one-on-one 1.5 hour interviews with
 6-7 student volunteers for each tutorial using a think-aloud protocol.~\cite{chi} In this protocol, students are asked
to work on a tutorial while talking aloud so that we could follow their thought processes. Hints are provided
as appropriate. These individual interviews provide an opportunity to probe students' thinking as they work through
a tutorial and gauge the extent to which students are able to benefit from them. After each of these interviews, 
the tutorials are modified based upon the feedback obtained. 
Then, they are administered in the classroom and are modified further based upon the feedback. Table 1 shows the
performance on the pre-/post-test of advanced undergraduate students in a quantum mechanics course on the last version.
The pre-test was given after traditional instruction on relevant concepts but before the tutorial.
Below we summarize each tutorial and discuss student performance on the case-study. We note that the pre-test
and post-test for a QuILT were not identical but often had some identical questions. 

\subsection{Time-development QuILT}

One difficulty with the time development of wave functions stems from the fact that many students believe that the only
possible wave functions for a system are stationary states.~\cite{phystoday,quilt} Since the Hamiltonian
of a system governs its time development, we may
expand a non-stationary state wave function $\Psi(x,0)$ at the initial time $t=0$ in terms of the stationary states and then multiply
appropriate time dependent phase factors $e^{-iE_n t/\hbar}$ with each term (they are in general
different for different stationary states because the energies $E_n$ are
different) to find the wave function $\Psi(x,t)$ at time $t$. Students often append an overall
time-dependent phase factor even if the wave function is in a linear superposition of
the stationary states.~\cite{phystoday} To elicit this difficulty, the pretest of this QuILT begins by asking students about
the time dependence of a non-stationary state wave function for
an electron in a one-dimensional infinite square well. If the students choose an overall phase factor similar to that for a stationary
state, they are asked for the probability density, i.e., the absolute square of the wave function.
As noted above, when a non-stationary state is expanded in terms of stationary states, 
the probability density at time $t$, $\vert \Psi(x,t) \vert^2$, is generally non-stationary due to a different time-dependent
phase factor in each term.
If students incorrectly choose that the wave function is time-independent even for a non-stationary state, arguing that
overall phase factors cancel out, the tutorial asks them to watch the simulations for the time evolution of the probability densities.
 
Simulations for this QuILT are adapted from the Open Source Physics simulations developed by Belloni and Christian.~\cite{mario,pqp}
These simulations are highly effective in challenging students' beliefs. Students are often taken aback when they find that the 
probability density oscillates back and forth for the non-stationary state. Figure 1 shows snapshots adapted in QuILT from
an Open Source Physics simulation by Belloni and Christian for the probability density for
a non-stationary state wave function for a one-dimensional harmonic oscillator well.
In the actual simulation, students watch the probability density evolve in time.

When students observe that the probability 
density does not depend on time for the stationary-state wave function but depends on time for the non-stationary-state
wave function, they are challenged
to resolve the discrepancy between their initial prediction and observation. 
In our model, this is a good time to provide students guidance and feedback to help them build a robust knowledge structure. 
Students then work through the rest of the QuILT which provides appropriate support and 
helps solidify basic concepts related to time development.
Students respond to time development questions with stationary and non-stationary state wave functions in
problems involving different potential energies (e.g., harmonic oscillator, free-particle etc.) 
to reinforce concepts, and they receive timely feedback to build their knowledge structure. 
For each case, they check their calculations and predictions for the time-dependence of the probability density in each case 
with the simulations. 
Within an interactive environment, they learn that the Hamiltonian governs the time development of the system, and that 
the eigenstates of the Hamiltonian are special with regards to the time evolution of the system. They learn that not
all possible wave functions are stationary-state wave functions, and they learn about 
the difference between the time-independent and time-dependent Schroedinger equation. 

Table 1 shows that in the case study in which nine students took both the pre-/post-tests, the average student performance
improved from $53\%$ to $85\%$ after working on the QuILT. 
As discussed earlier, the most common difficulty on the pre-test was treating the time evolution of
non-stationary states as though those states were stationary states.
Moreover, two students who were absent on the day the pre-test
and tutorial were administered in the class but were present for the post-test in the following class obtained $30\%$ and
$0\%$ on the post-test respectively.  

\subsection{Uncertainty Principle QuILT}

The QuILT on the uncertainty principle contains three parts with increasing levels of sophistication. 
Depending upon the level of students, the instructors may choose to use only one or all parts.
The first part of this QuILT helps students understand that this fundamental
principle is due to the wave nature of particles. With the help of the de Broglie relation, the QuILT helps students understand that a 
sinusoidal extended wave has a well-defined wavelength and momentum but does not have a well-defined position. On the other hand, a wave 
pulse with a well defined position does not have a well defined wavelength or momentum. 

Students gain further insight into the 
uncertainty principle in the second part of the QuILT by Fourier transforming the position-space wave function and noticing how the 
spread of the position-space wave function affects its spread in momentum space. 
Computer simulations involving Fourier transforms are exploited in 
this part of the QuILT and students Fourier transform various position-space wave function with different spreads and check the
corresponding changes in the momentum-space wave function.
The third part of the QuILT helps students generalize the
uncertainty principle for position and momentum operators to any two incompatible observables whose corresponding operators do not 
commute. This part of the QuILT also helps students bridge this new treatment with students' earlier encounter with
uncertainty principle for position and momentum in the context of the spread of a wave function in position and momentum space. The
QuILT also helps students understand why a measurement of one observable immediately followed by the measurement of another 
incompatible observable does not guarantee a definite value for the second observable. 

Table 1 shows that the average performance of 12 students who took the last version of the QuILT
improved from $42\%$ to $83\%$ from pre-test to post-test. In a question that was common for both the pre-test and post-test,
students were asked to make a qualitative sketch of the absolute value of the Fourier transform of a delta function. They were
asked to explain their reasoning and label the axes appropriately. Only one student in the pre-test drew a correct diagram. In
the post-test, 10 out of 12 students were able to draw correct diagrams with labeled axes and explain why the Fourier transform
should be a constant extended over all space. Also, in the post-test, 10 out of 12 students were able to draw the Fourier transform of 
a Gaussian position space wave function and were able to discuss the relative changes in the spread of the position and the
corresponding momentum space wave functions. These were concepts they had explored using computer simulations while working
on the QuILT. Similar results were found in individual interviews conducted earlier with other students during the development
of the QuILT.

One of the questions on both the pre-/post-test of this tutorial was the following:\\
Consider the following statements: ``Uncertainty principle makes sense.
When the particle is moving fast, the position measurement has uncertainty
because you cannot determine the particle's position precisely..it is a blur....that's
exactly what we learn in quantum mechanics..if the particle has a large speed, the
position measurement cannot be very precise."
Explain why you agree or disagree with the statement.\\
Out of the 12 students who took both pre-/post-tests, 7 students provided incorrect responses
on the pre-test.
The following are examples of incorrect student responses on the pre-test:
\begin{enumerate}
\item {\it I agree...when P is high it is easy to determine, while x is difficult to determine. The opposite is
also true, when P is small it is difficult to determine, while x is easy to determine.}
\item {\it I agree because when a particle has a high velocity it is difficult to measure the position accurately}
\item {\it I agree because I know the uncertainty principle to be true.}
\item {\it agree. When a particle is moving fast, we cannot determine its position exactly-it resembles a wave-at
fast speed, its momentum can be better determined.}
\end{enumerate}
In comparison, one student provided incorrect response and one did not provide a clear reasoning on the post-test. 

\subsection{Mach-Zehnder Interferometer QuILT}

The goals of this QuILT are to review
the interference at a detector due to the superposition of light from the two paths of the interferometer.
The tutorial adapts a simulation developed by Albert Huber
(http://www.physik.uni-muenchen.de/didaktik/Computer/interfer/interfere.html) to help students learn the following
important quantum mechanical concepts:
\begin{itemize}
\item interference of a single photon with itself after it passes through the two paths of the MZ.
\item effect of placing detectors and polarizers in the path of the photon in the MZ.
\item how the information about the path along which a photon went (``which-path" information) destroys the interference
pattern. 
\end{itemize}
A screen shot from the simulation is shown in Figure 2.

Students were given the following information about the setup:
The basic schematic setup for the Mach-Zehnder interferometer (MZ) used in this QuILT is as follows (see Figure 3) with changes made later in
the tutorial, e.g., changes in the position of the beam splitters, incorporation of polarizers, detectors or a glass piece, 
to illustrate various concepts. All angles of incidence are $45^0$ with respect to the normal to the surface.
For simplicity, we will assume that light can only reflect from one of the two surfaces of the identical half-silvered mirrors (beam splitters)
$BS_1$ and $BS_2$ because of anti-reflection coatings. The detectors $D_1$ and $D_2$ are \underline{point} detectors
located symmetrically with respect to the other components of the MZ as shown.
The photons originate from a monochromatic coherent point source.
Assume that the light through both the $U$ and $L$ paths travels
the same distance in \underline{vacuum} to reach each detector. \\

In this QuILT, students first learn about the basics of phase changes that take place as light reflects or passes through
different beam splitters and mirrors in the MZ setup by drawing analogy with reflected or transmitted wave on a string 
with fixed or free boundary condition at one end. Then, students use the simulation to
learn that a single photon can interfere with itself and produce interference pattern after it passes through both paths of the MZ.
Students explore and learn using simulations that
``which-path" information is obtained by removing $BS_2$ or by placing detectors or polarizers in certain locations.
Later in the tutorial, point detector $D_1$ is replaced with a screen.

Table 1 shows that the average performance of 12 students who took the last version of the MZ QuILT
improved from $48\%$ to $83\%$ from pre-test to post-test. Moreover, all but one of the 12 students in the post-test obtained 
perfect scores on the following three questions (correct options (c), (b), and (b) respectively) that were similar (but not necessarily 
identical to) the kinds of setups they had explored using the simulation within the guided QuILT approach:

\begin{enumerate}
\item
If you insert polarizers 1 and 2 (one with a horizontal and the other with a $45^0$
transmission axis) as in the Figure 4, how does the interference pattern compare with the case when the two polarizers have orthogonal transmission axis?\\
(a) The interference pattern is identical to the case when polarizers 1 and 2 have orthogonal axes.\\
(b) The interference pattern vanishes when the transmission axes of polarizers 1 and 2 are horizontal and $45^0$.\\
(c) An interference pattern is observed, in contrast to the case
when polarizers 1 and 2 were orthogonal to each other.\\
(d) No photons reach the screen when the transmission axes of polarizers 1 and 2 are horizontal and $45^0$.\\

\item
If you insert polarizer 1 with a horizontal transmission
axis and polarizer 2 (between the second
beam splitter and the screen) with a $45^0$ transmission axis (Figure 5), how does the interference pattern compare with the case when
only polarizer 1 was present?\\
(a) The interference pattern is identical to the case when only polarizer 1 was present.\\
(b) The intensity of the interference pattern changes but the interference pattern is maintained in the presence of polarizer 2.\\
(c) The interference pattern vanishes when polarizer 2 is inserted but some photons reach the screen.\\
(d) An interference pattern reappears that was absent when only polarizer 1 was present.

\item
If you insert polarizer 2 with a $45^0$ transmission axis between the second beam splitter and the screen (Figure 6), how does the interference pattern compare with the case when polarizer 2 was not present?\\
(a) The interference pattern is unchanged regardless of the presence of polarizer 2 because all interference effects occur before
beam splitter 2.\\
(b) The intensity of the interference pattern decreases but the interference pattern is maintained even in the presence of polarizer 2.\\
(c) The intensity of the interference pattern increases in the presence of polarizer 2.\\
(d) The interference pattern vanishes when polarizer 2 is inserted but some photons reach the screen.

\end{enumerate}

\subsection{Survey about QuILTs}

A survey of 12 students whose pre-/post-test data is presented in Table 1 was given to assess the effectiveness
of QuILTs from students' perspective. 
Below we provide the questions and student responses:\\
\begin{itemize}
\item Please rate the tutorials for their overall effectiveness where 1 means totally ineffective and 5 means very effective.\\
In response to this question, no student chose 1 or 2, one student chose 3, one chose 3.5, three chose 4, one 4.5 and six chose 5.

\item How often did you complete the tutorial at home that you could not complete during the class?
(1) Never, (2) less than half the time, (3) often, (4) most of the time, (5) always.\\
In response to this question, no student chose (1), one student chose (2), two students chose (3), 6 chose (4), and 3 chose (5).

\item How often were the hints/solutions provided for the tutorials useful?
(1) Never, (2) less than half the time, (3) often, (4) most of the time, (5) always.\\
In response to this question, no student chose (1) or (2), 2 students chose (3), 5 chose (4) and 5 chose (5).

\item Is it more helpful to do the tutorials in class or would you prefer to do them as homework? Please explain the
advantages and disadvantages as you see it.\\
In response to this question, 10 students felt that doing them in class was more useful. 
The students who preferred doing them in class often noted that the tutorials focused on improving their conceptual
understanding which was best done via group discussion and hence in class. They appreciated the fact that any questions
they had could be discussed and they benefited from the reasoning provided by their peers and instructor.
The few students who preferred doing them at home felt that more time and effort will go into them if they did them
at home.

\item How frequently should the tutorials be administered in the class (e.g., every other class,
once a week, once every other week)? Explain your reasoning.\\
A majority of students liked having the tutorials once a week. This frequency was considered to be the best
by some students who felt that the concepts learned in the tutorials made it easier for them to understand the textbook 
and homework problems later in the week and integrate the material learned. Others felt that once a week was the best 
because tutorials helped them 
focus on concepts that were missed in lectures, book, and student/teacher conversation.

\item Do you prefer a multiple-choice or open-ended question format for the tutorial questions?
Explain your reasoning.\\
Students in general seemed to like the questions that were in the multiple-choice format but most of them also
appreciated the open-ended questions. Some students noted that the multiple-choice questions helped focus their attention
on important issues and common difficulties and misconceptions while the open-ended questions stimulated creative thought.
Some students felt that multiple-choice format may be better for the ``warm-up" tutorial done at home and the open-ended 
questions may be better for the main tutorial done in the class. Some students felt that a mix of the two types of questions was
best because the multiple-choice format was a good way to get the fundamental concepts across and the open-ended questions gave
them an opportunity to apply these concepts and deepen their understanding of the concepts.

\end{itemize}

\section{Summary}

We have given an overview of the development of QuILTs and discuss the preliminary evaluation of three QuILTs
using pre-/post-tests in the natural classroom setting.
QuILTs adapt visualization tools to help students build physical intuition about quantum processes.
They are designed to help undergraduates sort through challenging quantum mechanics concepts.
They target misconceptions and common difficulties explicitly,
focus on helping students integrate qualitative and quantitative understanding, 
and learn to discriminate between concepts that are often confused.
They strive to help students develop the ability to apply quantum principles in different
situations, explore differences between classical and quantum ideas, and organize knowledge hierarchically.
Their development is an iterative process.
During the development of the existing QuILTs, we have conducted more than 100 hours of 
interviews with individual students to assess the aspects of the QuILTs that work well and those that require refinement.
QuILTs naturally lend themselves to dissemination via the web.
They provide appropriate feedback to students based upon their need and
are suited as an on-line learning tool for both undergraduates (and beginning graduate students)
in addition to being suitable  as supplements to lectures for a one or two-semester undergraduate quantum mechanics courses.

\section{Acknowledgments}

We are very grateful to Mario Belloni and Wolfgang Christian for their help in developing and adapting
their Open Source Physics simulations for QuILTs. We also thank 
Albert Huber for Mach Zehnder interferometer simulation and to Klaus Muthsam for the double slit simulation.
We thank all the faculty who have administered different versions of QuILTs in their classrooms.\\


\bibliographystyle{aipproc}

\pagebreak

\begin{table}[h]
\centering
\begin{tabular}[t]{|c|c|c|c|}
\hline
Tutorial& Number of students& $\%$ Pre-test Score&$\%$ Post-test Score \\[0.5 ex]
\hline \hline
Time development of wave function& 9&53&85\\[0.5 ex]
\hline
Uncertainty Principle&12&42&83\\[0.5 ex]
\hline
Mach-Zehnder Interferometer&12&48&83\\[0.5 ex]
\hline
\hline
\end{tabular}
\caption{Pre-/post-test performance of undergraduates in advanced quantum mechanics at the University of Pittsburgh
on the latest version of the three tutorials discussed.} \label{table1}
\end{table}




\begin{figure}
\includegraphics[height=.3\textheight]{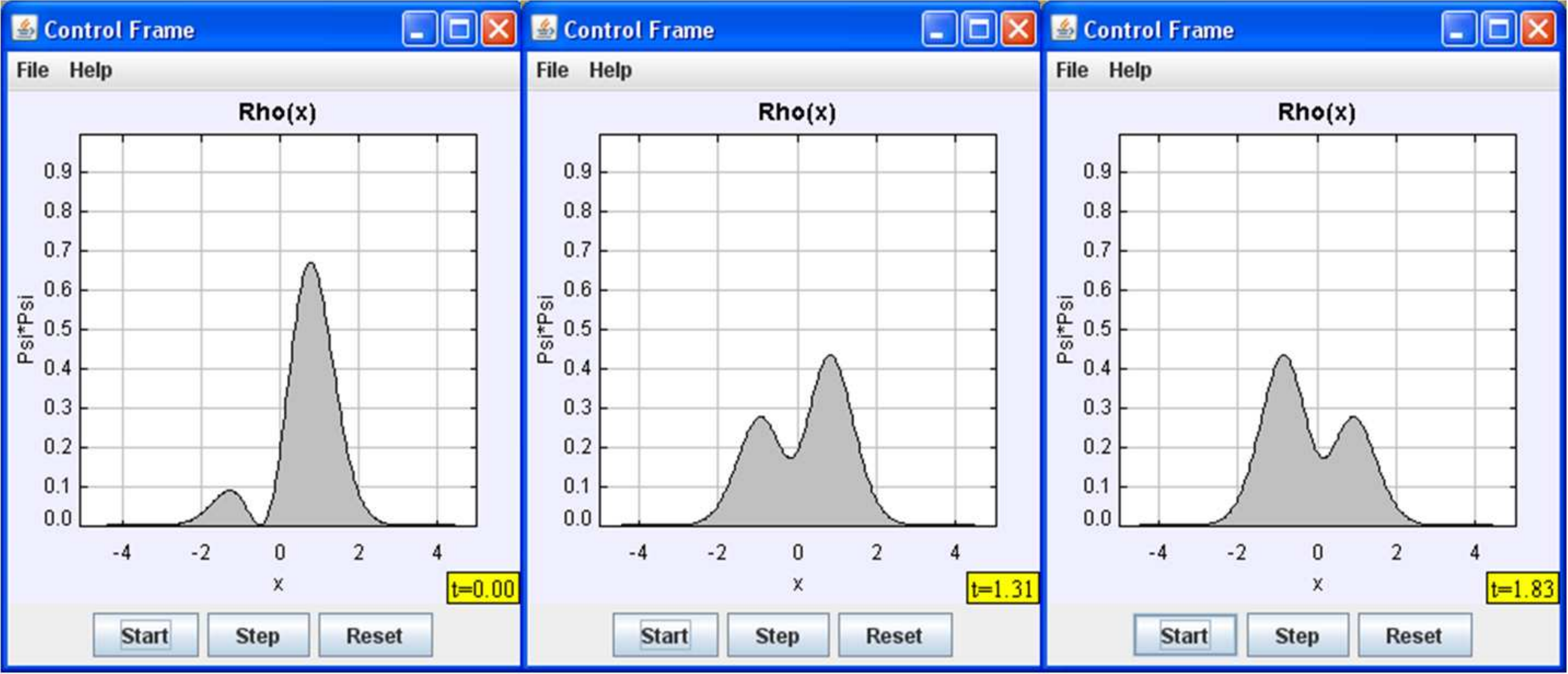}
\caption{
The snapshots at three different times of the probability density for a non-stationary state wave function (for
one-dimensional harmonic oscillator well) from the Open Source Physics simulations by Belloni
and Christian~\cite{mario}.}
\end{figure}


\begin{figure}
\includegraphics[height=.3\textheight]{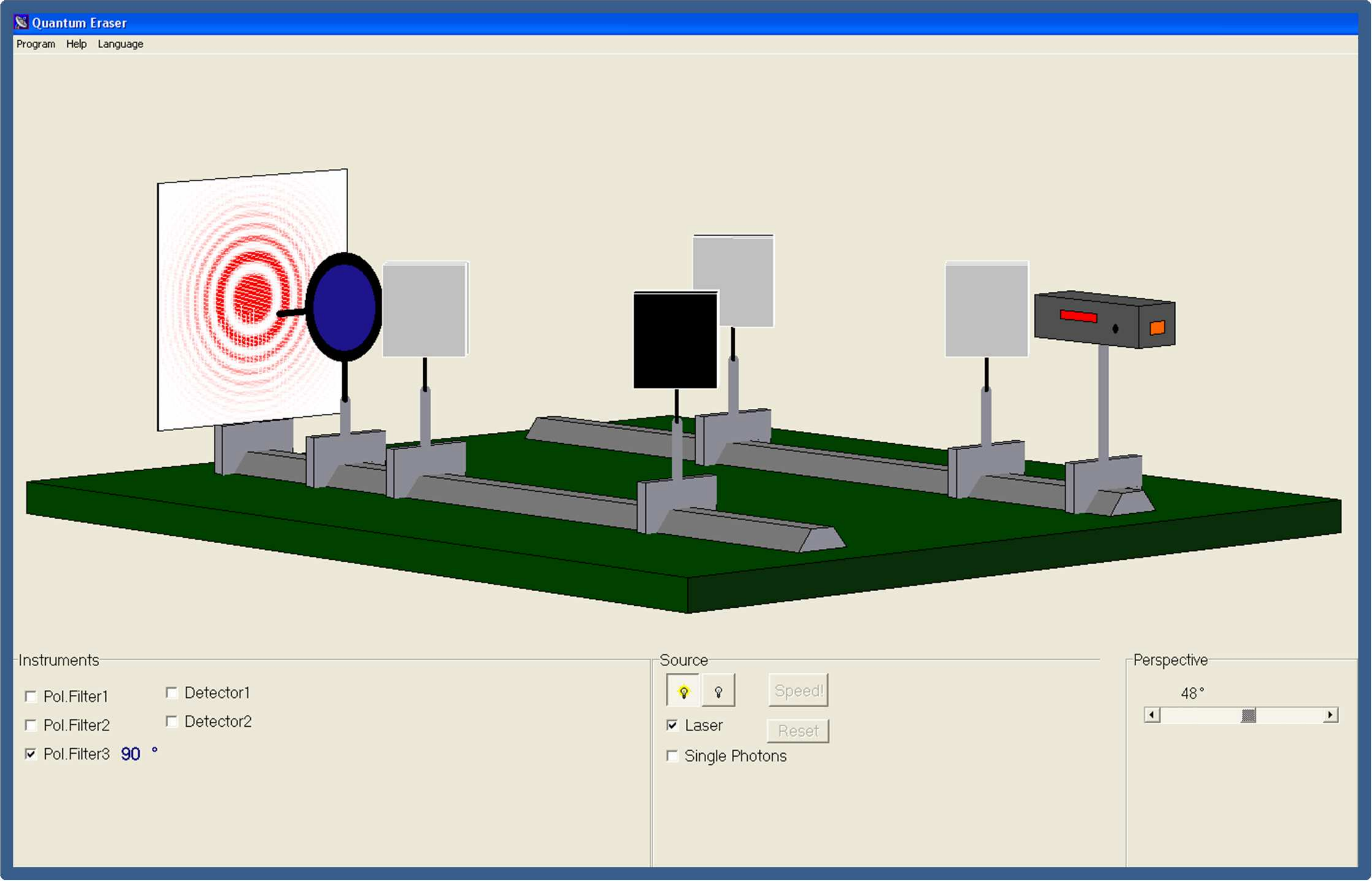}
\caption{
A screen shot of the setup of the Mach-Zehnder Interferometer simulation
by Huber~\cite{others} showing a polarizer right before the screen.
}
\end{figure}


\begin{figure}
\includegraphics[height=.3\textheight]{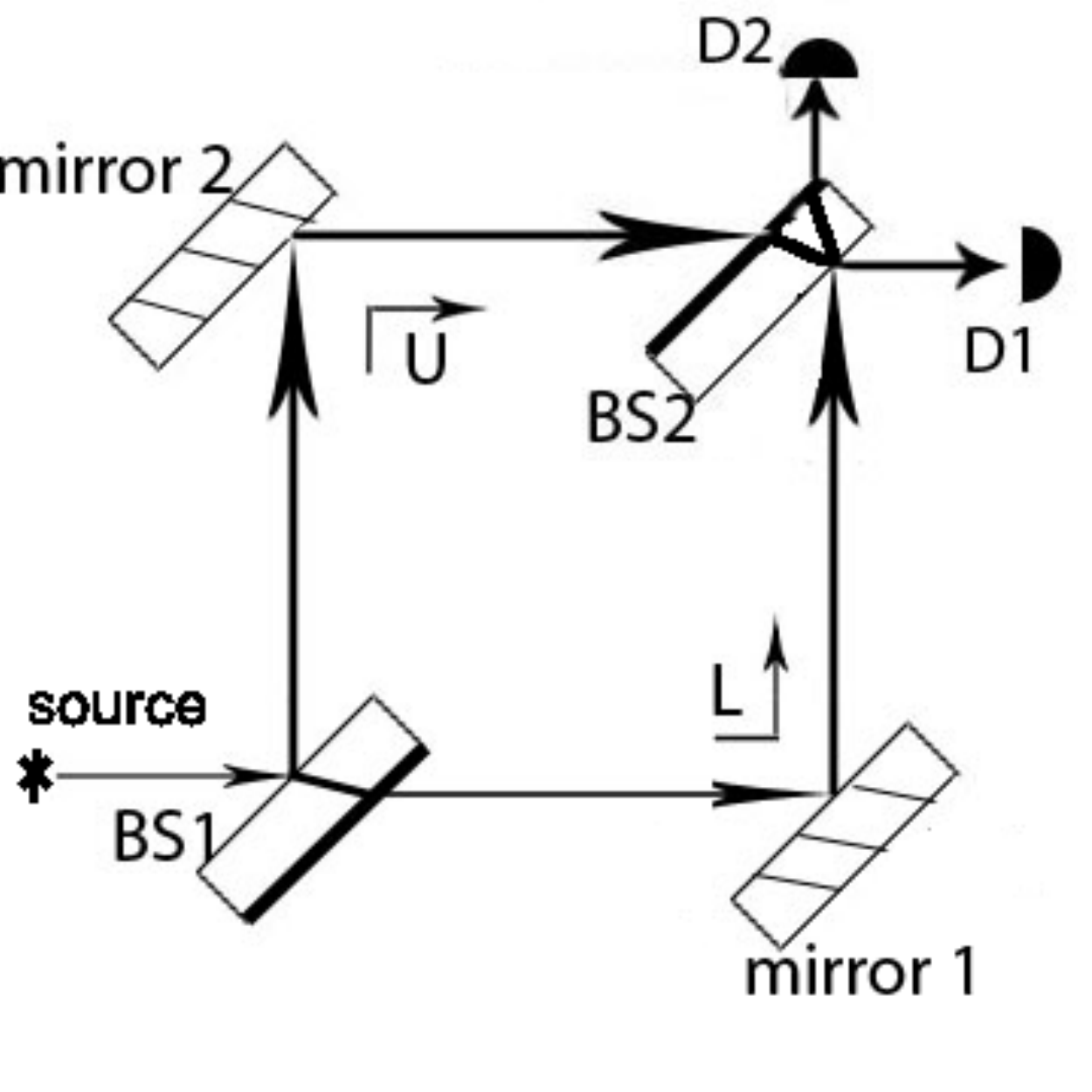}
\caption{
A schematic diagram of the setup in the Mach-Zehnder Interferometer simulation
by Huber~\cite{others}.
}

\end{figure}


\begin{figure}
\includegraphics[height=.3\textheight]{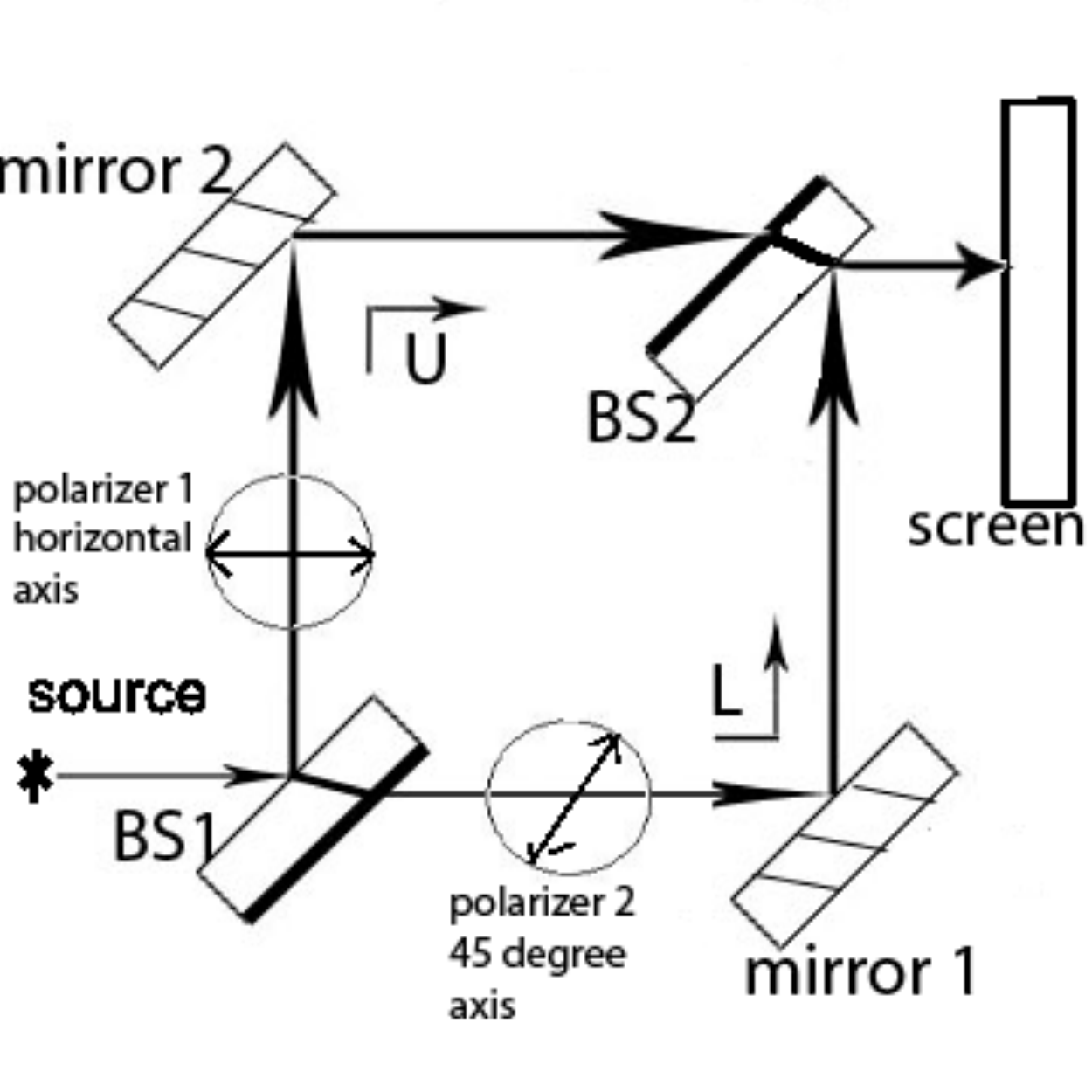}
\caption{
A schematic diagram of a modified setup with two polarizers in the Mach-Zehnder Interferometer simulation
by Huber~\cite{others}.}

\end{figure}


\pagebreak

\begin{figure}
\includegraphics[height=.3\textheight]{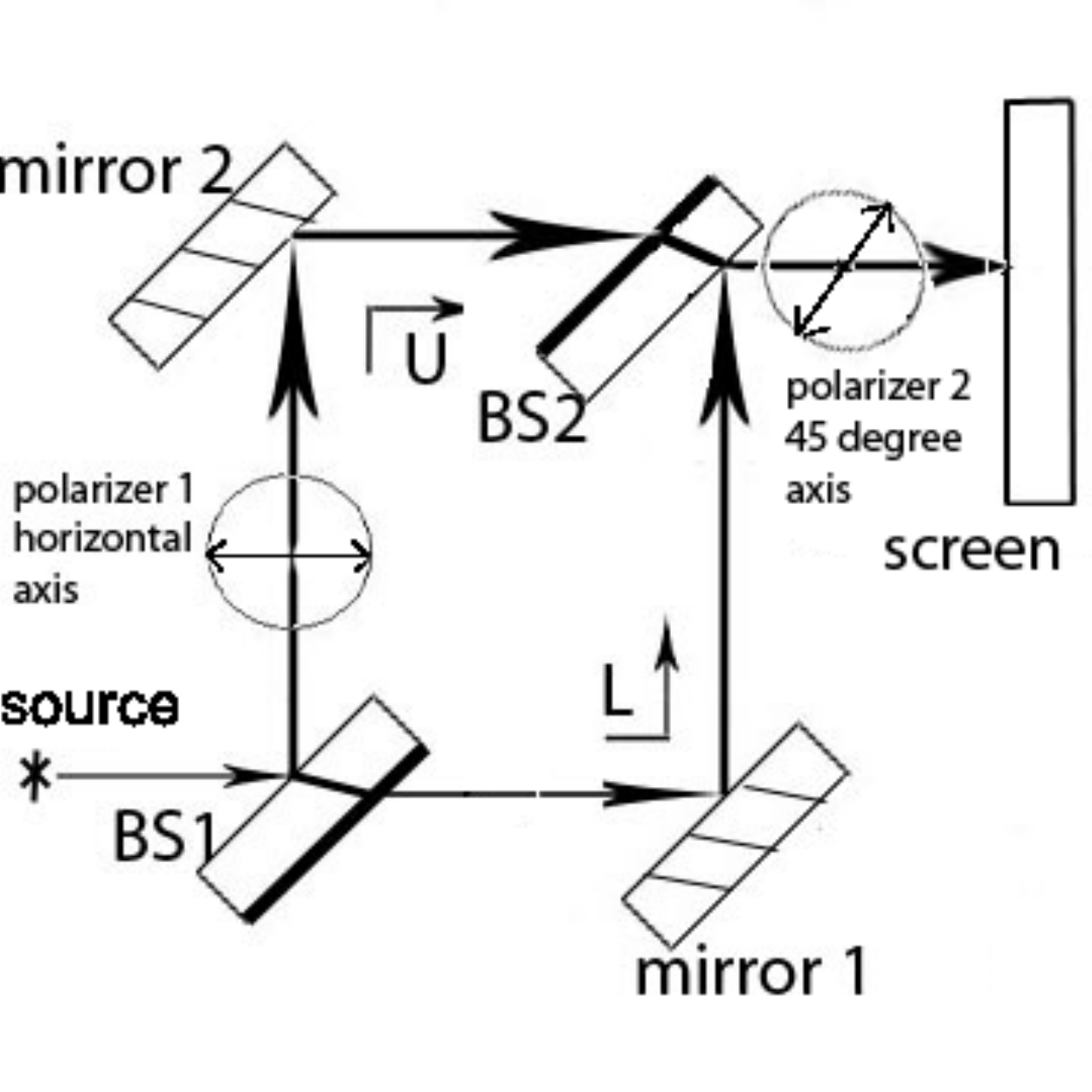}
\caption{
A schematic diagram of a modified setup with two polarizers in the Mach-Zehnder Interferometer simulation
by Huber~\cite{others}.
}

\end{figure}


\begin{figure}
\includegraphics[height=.3\textheight]{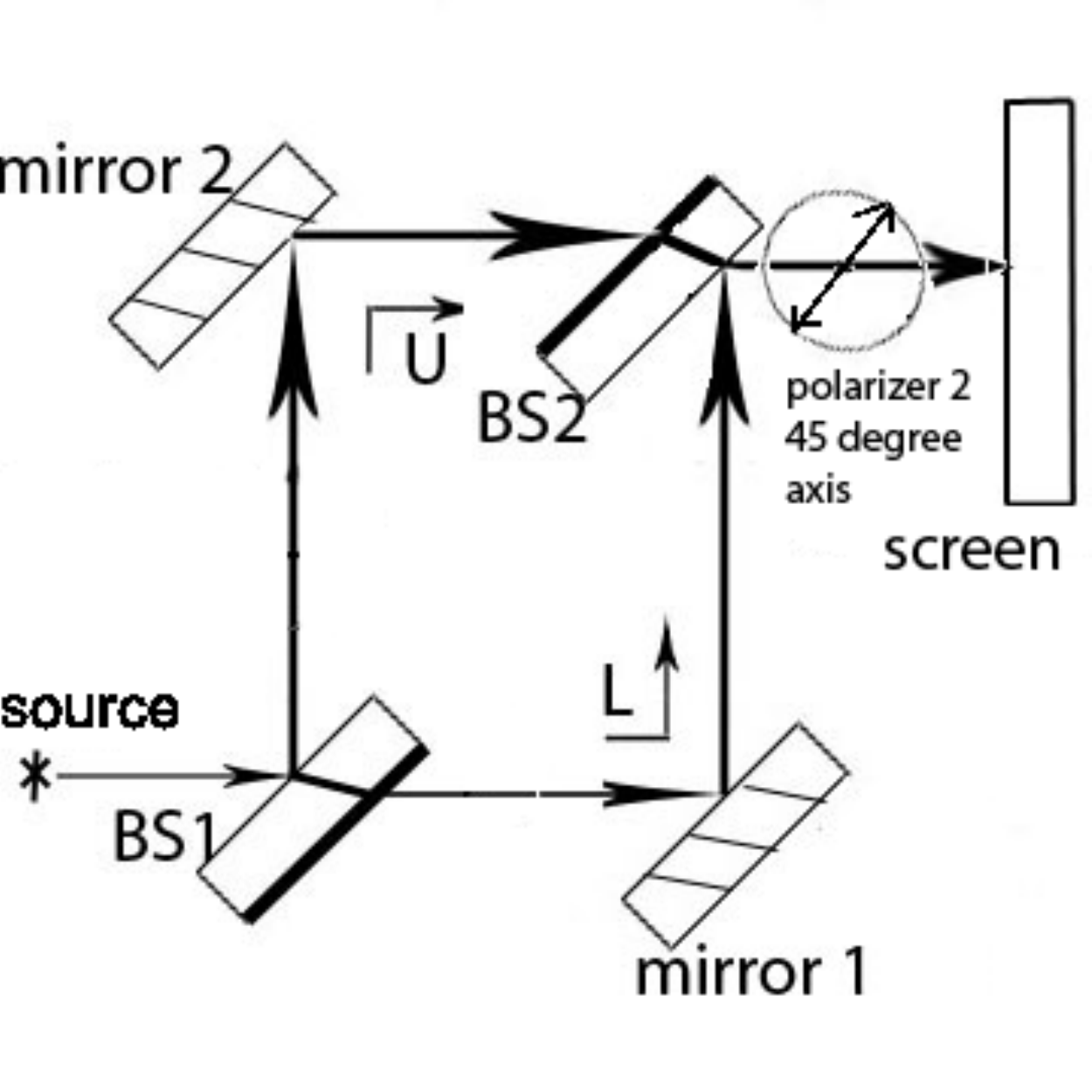}
\caption{
A schematic diagram of a modified setup with a polarizer after the second beam-splitter
in the Mach-Zehnder Interferometer simulation by Huber~\cite{others}.
}

\end{figure}

\end{document}